# Can error in quantum deletion machines be beneficial for deletion?

Sujan Vijayaraj [1], S. Balakrishnan [2]

**Abstract**

In this paper, a generalized input state dependent deletion machine called probabilistic quantum deletion machine is proposed. Considering the Pati-Braunstein deletion machine as a benchmark, the machine is characterized by its deletion probability and the probability of error in deletion. It is observed that there are parameters for which increase in error in deletion can increase the fidelity of deletion. It is also shown that the Pati-Braunstein deletion machine is easily derivable for a certain value of deletion probability. Further the deletion machine can offer better fidelity of deletion than the Pati-Braunstein deletion machine for any input state with the right parameters.

1. Introduction

The quantum no-deleting theorem proposed by Pati and Braunstein is an important no-go theorem in quantum information [1]. It states that an arbitrary quantum state cannot be deleted perfectly against a copy. This fundamental limitation occurs as a result of linearity in quantum mechanics. The theorem was quickly followed by proposals on approximate quantum deletion machines that have a deletion fidelity less than unity. This also includes a deletion machine from the authors of the theorem themselves [2] using the same operations stated in [1]. A good review of the early proposals of deletion machines is available in [3]. Note that the deletion operation should not be confused with erasure, which involves spending some energy [4].

A certain kind of generalized deletion machine has already been reported by Chakrabarty and Adhikari [5]. But the action of the machine stated in this paper is quite different from the machine proposed in [5]. The probabilistic quantum deletion machine that is introduced in our work is based

Sujan Vijayaraj
sujankvr@gmail.com

S. Balakrishnan
physicsbalki@gmail.com

1 School of Electronics Engineering, Vellore Institute of Technology, Vellore, India
2 Department of Physics, School of Advanced Sciences, Vellore Institute of Technology, Vellore, India

on the idea that not all copies get deleted due to its action but some are retained as it is. The operations also allow the machine to be interpreted as a Pati-Braunstein deletion machine [2] with an error. In spite of this, it is shown that any input state can undergo better deletion than the Pati-Braunstein deletion machine.

Quantum deletion in itself posits some weird possibilities. For instance, it was proven that the entangled output state of the Pati-Braunstein deletion machine does not violate Bell's inequality [5] but can still act as a teleportation channel [6]. A counterintuitive observation has also been made in this work, where under specific conditions, any increase in deletion probability in the machine decreases the deletion. Alternatively, any increase in the error in deletion, increases the fidelity of deletion sometimes even up to unit fidelity. This observation throws light on the weird nature of quantum deletion. It is also shown that partial information on the input states helps to determine the necessary parameters for more efficient deletion. This is highlighted using the range of values of $ab$, which is the product of the input state coefficients.

The paper is organized as follows. Firstly the Pati-Braunstein deletion machine is described and its corresponding fidelities are calculated. Then our probabilistic deletion machine is introduced and the density matrix corresponding to different modes is derived. The fidelities are calculated and using which the proposed deletion machine is analyzed for different sets of parameters. Finally, our observations essentially indicate that the proposed quantum deletion machine is vastly different from the existing deletion machines.

## 2. Pati-Braunstein deletion machine

The action of the Pati-Braunstein (P-B) deletion machine is defined as follows,

$$|0\rangle|0\rangle|A\rangle \rightarrow |0\rangle|\Sigma\rangle|A_0\rangle$$
$$|0\rangle|1\rangle|A\rangle \rightarrow |0\rangle|1\rangle|A\rangle$$
$$|1\rangle|0\rangle|A\rangle \rightarrow |1\rangle|0\rangle|A\rangle$$
$$|1\rangle|1\rangle|A\rangle \rightarrow |1\rangle|\Sigma\rangle|A_1\rangle$$

(1)

where $|\Sigma\rangle = M_0|0\rangle + M_1|1\rangle$ is the blank state, $|A\rangle$ is the initial ancilla state and $|A_0\rangle$, $|A_1\rangle$ are the final ancilla states. For the input state $|\Psi\rangle = a|0\rangle + b|1\rangle$ (where $a$ and $b$ are unknown complex numbers such that $|a|^2 + |b|^2 = 1$),

$$|\Psi\rangle|\Psi\rangle|A\rangle = [a^2|0\rangle|0\rangle + b^2|1\rangle|1\rangle + ab(|0\rangle|1\rangle + |1\rangle|0\rangle)]|A\rangle$$
$$= a^2|0\rangle|\Sigma\rangle|A_0\rangle + b^2|1\rangle|\Sigma\rangle|A_1\rangle + ab(|0\rangle|1\rangle + |1\rangle|0\rangle)|A\rangle. \quad (2)$$

The reduced density matrix for the output state in mode 1 is given by,

$$\rho_1 = |a|^4|0\rangle\langle 0| + |b|^4|1\rangle\langle 1| + |a|^2|b|^2 I.$$

(3)

The fidelity of the qubit in mode 1, $F_1 = \langle \Psi|\rho_1|\Psi\rangle = 1 - 2|a|^2|b|^2$. This gives a measure of the retention of the qubit. Similarly the reduced density matrix for the output state in mode 2 is given by,

$$\rho_2 = (|a|^4 + |b|^4)|\Sigma\rangle\langle\Sigma| + |a|^2|b|^2 I.$$

(4)

The fidelity of the qubit in mode 2, known as the fidelity of deletion, $F_2 = \langle\Sigma|\rho_2|\Sigma\rangle = 1 - |a|^2|b|^2$. The plot in Fig. 1 shows the relation between $F_2$ and $ab$. It is simple to show that $F_2 - F_1 = |a|^2|b|^2$. This clearly indicates that the difference in the fidelities cannot be zero for the arbitrary input states.

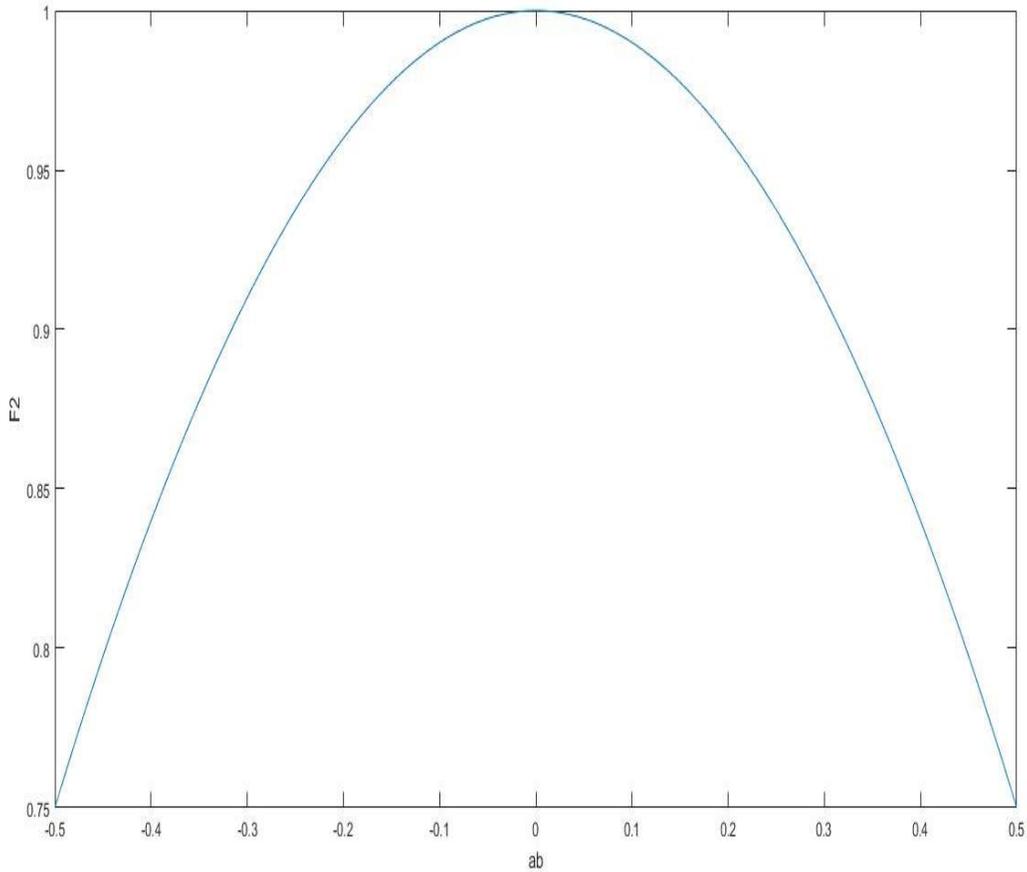

**Fig. 1** Relation between $ab$ and deletion fidelity $F_2$ for P-B machine

### 3. Probabilistic quantum deletion machine

The action of the probabilistic deletion machine is as follows,

$$|0\rangle|0\rangle|A\rangle \rightarrow p|0\rangle|\Sigma\rangle|A_0\rangle + q|0\rangle|0\rangle|A\rangle$$
$$|0\rangle|1\rangle|A\rangle \rightarrow |0\rangle|1\rangle|A\rangle$$
$$|1\rangle|0\rangle|A\rangle \rightarrow |1\rangle|0\rangle|A\rangle$$
$$|1\rangle|1\rangle|A\rangle \rightarrow p|1\rangle|\Sigma\rangle|A_1\rangle + q|1\rangle|1\rangle|A\rangle$$

(5)

where $p$, $q$ are complex numbers such that $|p|^2 + |q|^2 = 1$ and $p \neq 0$. Notice that our deletion machine includes a kind of error in deletion to the Pati-Braunstein deletion machine. The output state undergoes deletion, but with a probability of $|p|^2$ which we will call the deletion probability. Rest of the time, the state remains unchanged with a probability of $|q|^2$ which is the probability of error in deletion. The proposed machine is much broader than the Pati-Braunstein deletion machine as it reasonably accommodates error in the process of deletion. Note that when there is no error ($q = 0$), the machine condenses to the Pati-Braunstein deletion machine.

Considering the state $|\Psi\rangle = a|0\rangle + b|1\rangle$ as the input to the machine, we have

$$|\Psi\rangle|\Psi\rangle|A\rangle = [a^2|0\rangle|0\rangle + b^2|1\rangle|1\rangle + ab(|0\rangle|1\rangle + |1\rangle|0\rangle)] |A\rangle$$
$$= pa^2|0\rangle|\Sigma\rangle|A_0\rangle + qa^2|0\rangle|0\rangle|A\rangle + pb^2|1\rangle|\Sigma\rangle|A_1\rangle +$$
$$qb^2|1\rangle|1\rangle|A\rangle + ab(|0\rangle|1\rangle + |1\rangle|0\rangle)|A\rangle.$$

(6)

The reduced density matrix for the output state in mode 1 is given by,

$$\rho_1 = (|a|^4|p|^2 + |a|^4|q|^2 + |a|^2|b|^2)|0\rangle\langle 0| + (qab^*|a|^2 + q^*ab^*|b|^2)|0\rangle\langle 1| + (qa^*b|b|^2 +$$
$$q^*a^*b|a|^2)|1\rangle\langle 0| + (|b|^4|p|^2 + |b|^4|q|^2 + |a|^2|b|^2)|1\rangle\langle 1|$$
$$= |a|^2|0\rangle\langle 0| + |b|^2|1\rangle\langle 1| + (qab^*|a|^2 + q^*ab^*|b|^2)|0\rangle\langle 1| + (qa^*b|b|^2 +$$
$$q^*a^*b|a|^2)|1\rangle\langle 0|.$$

(7)

The fidelity of the qubit in mode 1,

$$F_1 = \langle\Psi|\rho_1|\Psi\rangle = 1 - (2 + q + q^*) |a|^2|b|^2.$$

(8)

The reduced density matrix for the output state in mode 2 is given by,

$$\begin{aligned}
\rho_2 &= (|a|^4|p|^2 + |b|^4|p|^2)|\Sigma\rangle\langle\Sigma| + (|a|^4|q|^2 + |a|^2|b|^2)|0\rangle\langle 0| + (qab^*|a|^2 + \\
&\quad q^*ab^*|b|^2)|0\rangle\langle 1| + (qa^*b|b|^2 + q^*a^*b|a|^2)|1\rangle\langle 0| + (|b|^4|q|^2 + |a|^2|b|^2)|1\rangle\langle 1| \\
&= |p|^2(1 - 2|a|^2|b|^2)|\Sigma\rangle\langle\Sigma| + |a|^4|q|^2|0\rangle\langle 0| + (qab^*|a|^2 + q^*ab^*|b|^2)|0\rangle\langle 1| + \\
&\quad (qa^*b|b|^2 + q^*a^*b|a|^2)|1\rangle\langle 0| + |b|^4|q|^2|1\rangle\langle 1| + |a|^2|b|^2 I.
\end{aligned} \qquad (9)$$

The fidelity of the qubit in mode 2,

$$F_2 = \langle\Sigma|\rho_2|\Sigma\rangle = |p|^2(1 - 2|a|^2|b|^2) + |q|^2(|a|^4|M_0|^2 + |b|^4|M_1|^2) + |a|^2|b|^2 + ab^*M_0^*M_1(q|a|^2 + q^*|b|^2) + a^*bM_0M_1^*(q^*|a|^2 + q|b|^2). \qquad (10)$$

Let $|\Sigma\rangle = |+\rangle$, therefore $M_0 = M_1 = \frac{1}{\sqrt{2}}$ and,

$$F_2 = (1 - 0.5|q|^2)(1 - 2|a|^2|b|^2) + |a|^2|b|^2 + 0.5[ab^*(q|a|^2 + q^*|b|^2) + a^*b(q^*|a|^2 + q|b|^2)]. \qquad (11)$$

Thus we have obtained the expressions for the fidelity of retention (Eq. (8)) as well as for the fidelity of deletion (Eq. (11)) and they will be used to characterize the proposed deletion machine.

### 4. Analyzing fidelities

Let $|\Psi\rangle = |+\rangle$ and $p, q$ be non-negative real numbers. The fidelity of retention from Eq. (8) thereby reduces to

$$F_1 = 0.5(1 - q), \qquad (12)$$

and the fidelity of deletion from (11) reduces to

$$F_2 = -0.25q^2 + 0.5q + 0.75. \qquad (13)$$

Figure 2 shows the plot between $p$ and the reduced form of the fidelities $F_1$ and $F_2$. Note that $p$ and $q$ will be used interchangeably following the condition, $|p|^2 + |q|^2 = 1$. For the given input state, the fidelity of deletion gets closer to one as $p$ tends towards zero as seen from the figure. Note that $|\Psi\rangle = |+\rangle$ is the only input state for which there is no maximum for the fidelity of deletion because $p$ cannot be zero. The fidelity of retention increases with increase in $p$ up till $F_1 = 0.5$. However the fidelity of deletion decreases with increase in $p$ till $F_2 = 0.75$. Hence there are specific cases where increase in deletion probability actually decreases the deletion. Alternatively, it can be said that increase in the probability of error in deletion increases deletion fidelity. This is counterintuitive because increase in $p$ increases the number of deleted states, which should have always increased $F_2$. Also $F_2$ approaches 1 as $p$ tends to 0 for the described parameters in the deletion machine. This is the weird quality of the proposed quantum deletion machine. For the $|+\rangle$ state, difference in fidelities, $\Delta F = F_2 - F_1 = -0.25q^2 + q + 0.25$. This difference $\Delta F$ helps us to understand the variation between deletion and retention actions of the machine. Figure 3 shows the plot between $\Delta F$ and $p$. It is clear that the difference in fidelities decreases as $p$ increases. Observe that the difference in fidelities approaches the minimum value of $0.25$ for $p = 1$, which is nothing but the Pati-Braunstein deletion machine.

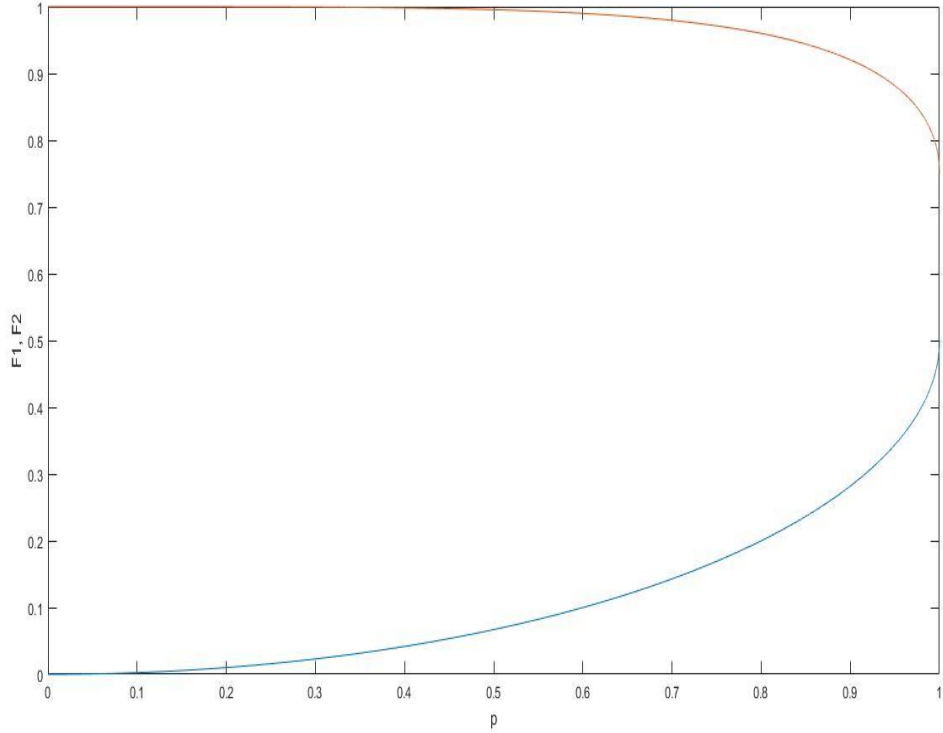

**Fig. 2** Relation between $p$ and $F_1$ (blue), $F_2$ (red) when $|\Psi\rangle = |+\rangle$

From Eq. (8) and Eq. (11), the fidelities assume the following forms when $a, b$ are real:

$$F_1 = 1 - 2a^2b^2(1 + q)$$

(14)

and

$$F_2 = (1 - 2a^2b^2)(1 - 0.5q^2) + qab + a^2b^2.$$

(15)

For $q = 0$ or $p = 1$, $F_1 = 1 - 2a^2b^2$ and $F_2 = 1 - a^2b^2$, which in fact is the case of the already known Pati-Braunstein deletion machine. Hence our machine describes a more general form from which the Pati-Braunstein deletion machine can be derived.

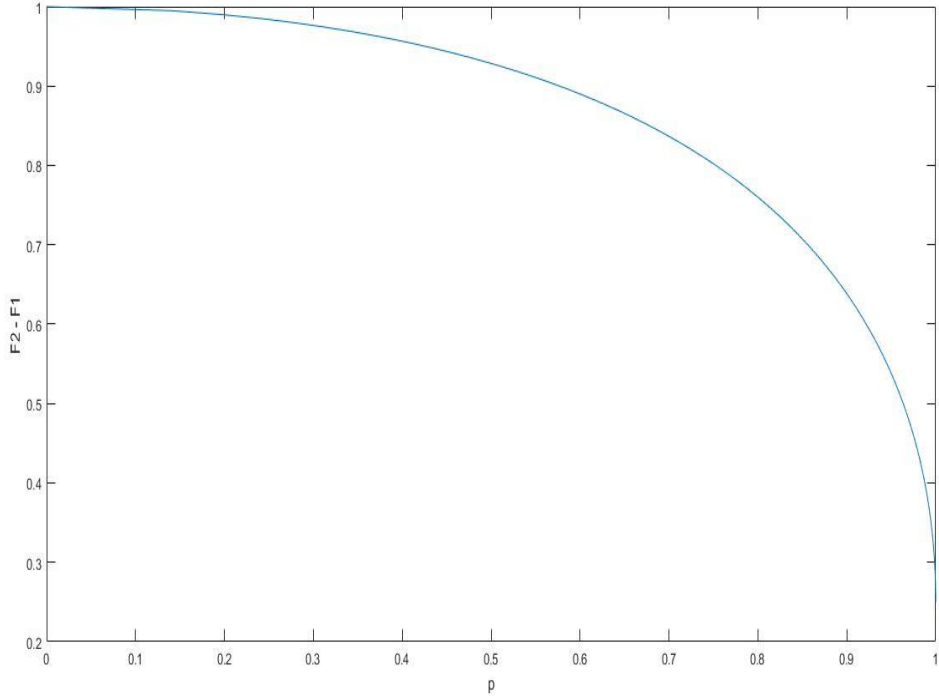

**Fig. 3** Relation between $p$ and $\Delta F$ when $|\Psi\rangle = |+\rangle$

The input state $|\Psi\rangle$, can also be different from the $|+\rangle$ state that was considered initially. For the initial state $|\Psi\rangle = \frac{\sqrt{3}}{2}|0\rangle + \frac{1}{2}|1\rangle$, the fidelities in Eq. (14) and eq. (15) reduce to $F_1 = 0.625 - 0.375q$ and $F_2 = -0.3125q^2 + 0.433q + 0.8125$. The plots of $F_1$ and $F_2$ with respect to $p$ is shown in Fig. 4. The fidelity of deletion, in this case, increases and then decreases near the end. It is reasonably high for most input states. On the other hand, fidelity of retention increases with $p$. The maximum fidelity of deletion is $0.975$. It is also evident that the maximum fidelity of deletion varies with respect to input states.

Assume $p = 0.5$, to obtain the relation between the fidelity of deletion and the input state for a fixed $p$. Thereby from Eq. (15), $F_2 = -0.25a^2b^2 + \sqrt{3}/2\ ab + 0.625$. The plot between this $F_2$ and $ab$ is shown in Fig. 5. By comparing this with Fig. 1, we observe that $F_2$ increases with increase in $ab$ till its maximum value of $0.9955$. The minimum value is $0.1295$. Hence the deletion machine allows drastically higher as well as lower values compared to the Pati-Braunstein deletion machine for different input states. Partial information about the input states can be helpful for narrowing down the parameters needed so as to offer more deletion. For example, when $p = 0.5$, deletion is much better when $ab > 0$. Higher the value of $ab$, more efficient the deletion machine when $p, q$ are greater than zero. This is shown in Table 1 for different values of $ab$. Notice how the minimum value of fidelity of deletion increases as $ab$ increases, making the deletion machine more efficient with reduced standard deviation in $F_2$. Also the minimum value of $F_2$ decreases as $p$ decreases as seen in Table 2. Hence the machine is better off with higher values of $p$ when the state is completely uncertain. As discussed above, the deletion machine can be seen as a Pati-Braunstein machine with some error when $p \neq 1$. Naturally one would expect reduced fidelities in such a machine. But on closer inspection, the deletion machine can offer much better fidelity of deletion, like when $ab > 0.336$ for $p = 0.5$.

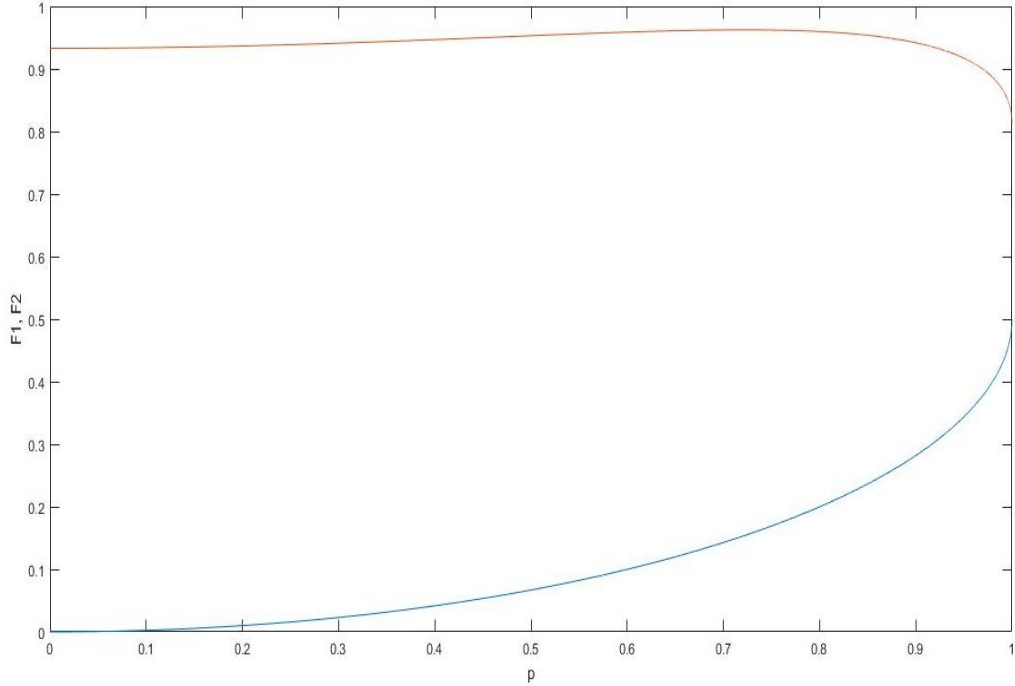

**Fig. 4** Relation between $p$ and $F_1$ (blue), $F_2$ (red) when $|\Psi\rangle = \frac{\sqrt{3}}{2}|0\rangle + \frac{1}{2}|1\rangle$

**Table 1** Relation between $ab$ and maximum, minimum, SD of $F_2$ from Eq. (15), where $\varepsilon \to 0$

| $ab$ | Minimum of $F_2$ | $p$ value for minimum $F_2$ | Maximum of $F_2$ | $p$ value for maximum $F_2$ | Standard deviation (SD) of $F_2$ |
|---|---|---|---|---|---|
| -0.25 | 0.2500 | $\varepsilon$ | 0.9375 | 1.000 | 0.1860 |
| -0.10 | 0.4000 | $\varepsilon$ | 0.9900 | 1.000 | 0.1684 |
| 0.10 | 0.6000 | $\varepsilon$ | 0.9951 | 0.995 | 0.1244 |
| 0.25 | 0.7500 | $\varepsilon$ | 0.9732 | 0.958 | 0.0763 |
| 0.30 | 0.8000 | $\varepsilon$ | 0.9649 | 0.931 | 0.0577 |
| 0.35 | 0.8500 | $\varepsilon$ | 0.9586 | 0.886 | 0.0384 |
| 0.40 | 0.8400 | 1 | 0.9576 | 0.809 | 0.0212 |
| 0.45 | 0.7975 | 1 | 0.9677 | 0.654 | 0.0211 |

The maximum fidelity is seen for the given input state when $q = ab/(1 - 2a^2b^2)$, where $a^2 \neq b^2 \neq 0.5$. This is obtained by equating the first derivative of the general form of $F_2$ in Eq. (15) with zero. Substituting $q$ back in $F_2$, we get $F_2 = 1 - a^2b^2 + a^2b^2/(1 - 2a^2b^2)$, which is the maximum possible fidelity of deletion for a given input state. The plot between the maximum value of $F_2$ and $a$ is shown in Fig. 6. As seen from Fig. 6, least value of the maximum fidelities of deletion among all the given input states is $0.9571$ (approximated up to 4 decimal places).

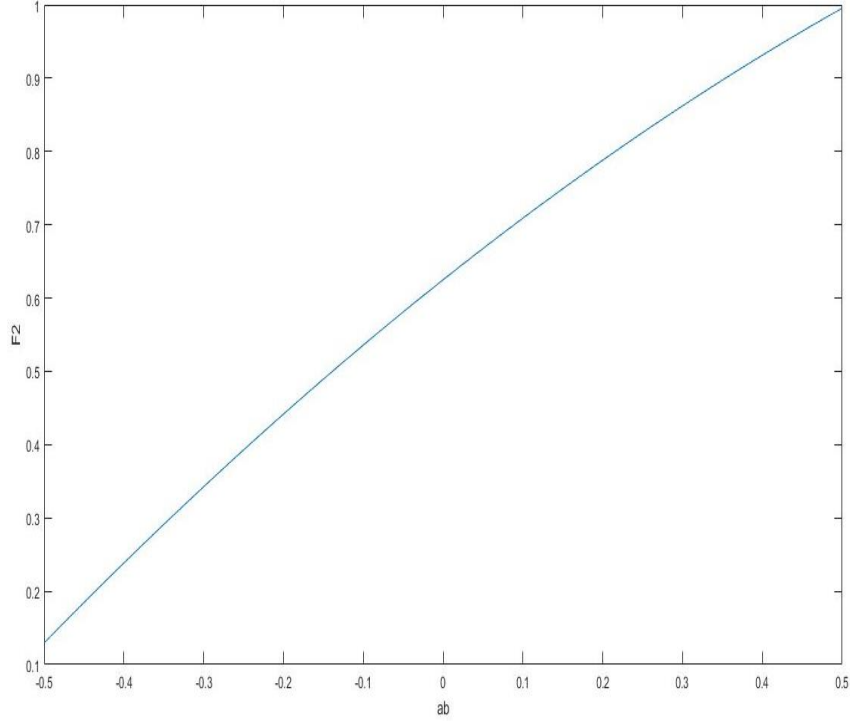

**Fig. 5** Relation between $ab$ and $F_2$ when $p = 0.5$

**Table 2** Relation between $p$ and the minimum, maximum of $F_2$ from Eq. (15), where $\varepsilon \to 0$

| $p$ | Minimum of $F_2$ | $ab$ value for minimum $F_2$ | Maximum of $F_2$ | $ab$ value for maximum $F_2$ |
|---|---|---|---|---|
| 0.250 | 0.0315 | $-0.5$ | 0.9970 | 0.5 |
| 0.500 | 0.1295 | $-0.5$ | 0.9955 | 0.5 |
| 0.750 | 0.3099 | $-0.5$ | 0.9713 | 0.5 |
| 0.900 | 0.4846 | $-0.5$ | 0.9636 | 0.2691 |
| 0.950 | 0.5695 | $-0.5$ | 0.9783 | 0.1730 |
| 0.990 | 0.6754 | $-0.5$ | 0.9951 | 0.0720 |
| 0.999 | 0.7271 | $-0.5$ | 0.9995 | 0.0224 |
| 1.000 | 0.7500 | $\pm 0.5$ | $1-\varepsilon$ | $\varepsilon$ |

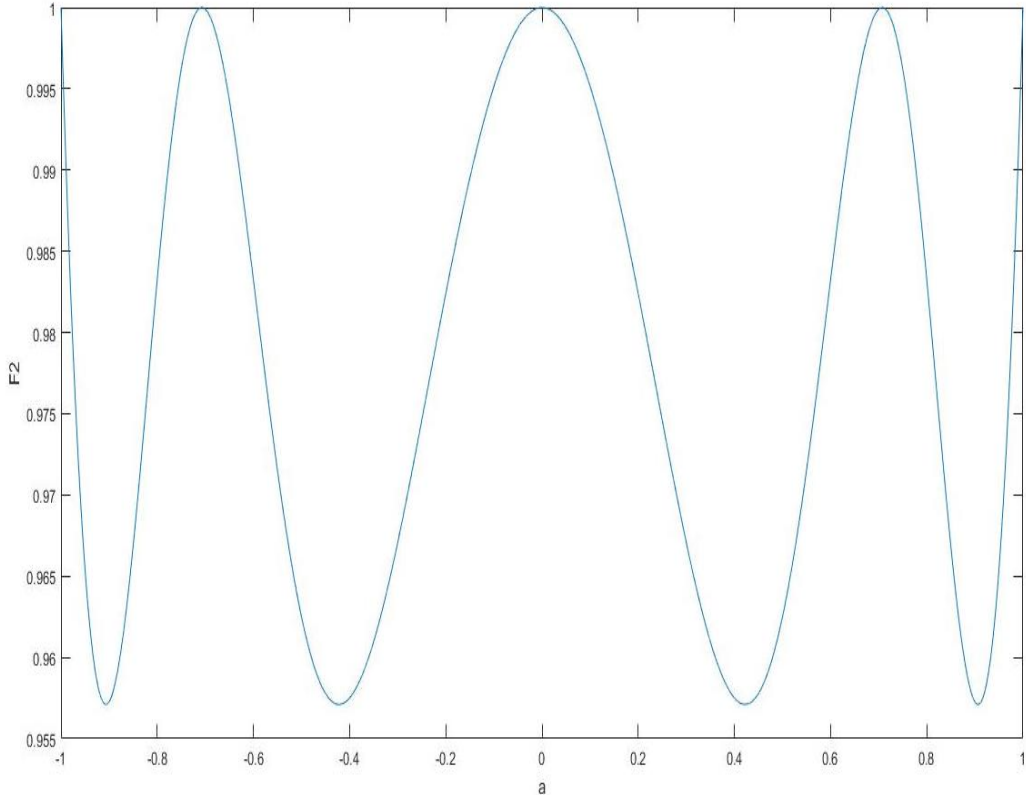

**Fig. 6** Relation between $a$ and corresponding maximum $F_2$

## 5. Conclusion

Perfect deletion against an unknown copy is impossible with quantum information. Even for approximate deletion, the fidelity of deletion can be low in many existing machines. Our probabilistic deletion machine on the other hand encompasses a wide range of fidelities that have been unexplored. Their variations with different input states and other parameters have also been discussed. One important observation is that the fidelity of deletion can decrease with increase in probability of deletion. This is to say that an error in quantum deletion can be beneficial. The machine also reduces to the Pati-Braunstein deletion machine when the probability of deletion is one. The deletion machine also offers better deletion than the Pati-Braunstein deletion machine for certain input states. Partial information about the input states can also help us to narrow down the parameters needed in the machine for better deletion. The maximum possible fidelity of deletion for all input states is always greater than $0.957$.

**References**

1. Pati, A. K., & Braunstein, S. L. (2000). Impossibility of deleting an unknown quantum state. *Nature*, *404*(6774), 164-165.

2. Pati, A. K., & Braunstein, S. L. (2000). Quantum no-deleting principle and some of its implications. *arXiv preprint quant-ph/0007121*.

3. Adhikari, S. (2009). Quantum Cloning and Deletion in Quantum Information Theory. *arXiv preprint arXiv:0902.1622*.

4. Landauer, R. (1961). Irreversibility and heat generation in the computing process. *IBM journal of research and development*, *5*(3), 183-191.

5. Chakrabarty, I., & Adhikari, S. (2005). A Generalized Deletion Machine. *arXiv preprint quant-ph/0511211*.

6. Bell, J.S.: On the Einstein-Podolsky-Rosen Paradox. Physics **1**, 195 (1964)

7. Chakrabarty, I., Ganguly, N., & Choudhury, B. S. (2011). Deletion, Bell's inequality, teleportation. *Quantum Information Processing*, *10*(1), 27-32.